\documentclass[aps,pra,superscriptaddress,12pt,tightenlines,nofootinbib]{revtex4}
\usepackage{amsmath,amsthm,graphicx,amssymb,verbatim,listings}
\usepackage{wasysym}
\usepackage[T1]{fontenc}     % needed for the guillemets
\usepackage{lmodern}         % needed to fix suboptimal font rendering
\usepackage[ pdftex, plainpages = false, pdfpagelabels,
                 pdfpagelayout = useoutlines,
                 bookmarks,
                 bookmarksopen = true,
                 bookmarksnumbered = true,
                 breaklinks = true,
                 linktocpage=all,
                 pagebackref=false,
                 colorlinks = true,
                 linkcolor = BrickRed,
                 urlcolor  = blue,
                 citecolor = BrickRed,
                 anchorcolor = green,
                 hyperindex = true,
                 hyperfigures
                 ]{hyperref}
\usepackage[usenames, dvipsnames]{xcolor}
\usepackage{xifthen} % provides \isempty test

\newcommand{\arxiv}[2][]{\ifthenelse{\isempty{#1}}{\href{http://arxiv.org/abs/#2}{{\tt arXiv:\allowbreak{}#2}}} {\href{http://arxiv.org/abs/#2}{{\tt arXiv:\allowbreak{}#2 [#1]}}}}

\begin{document}

\title{Maximal Sets of Equiangular Lines}

\author{Blake C.\ Stacey}
\affiliation{\href{http://www.physics.umb.edu/Research/QBism/}{QBism Research Group}, University of Massachusetts Boston}

\date{\today}

\begin{abstract}
  I introduce the problem of finding maximal sets of equiangular
  lines, in both its real and complex versions, attempting to write
  the treatment that I would have wanted when I first encountered the
  subject. Equiangular lines intersect in the overlap region of
  quantum information theory, the octonions and Hilbert's twelfth
  problem.  The question of how many equiangular lines can fit into a
  space of a given dimension is easy to pose --- a high-school student
  can grasp it --- yet it is hard to answer, being as yet unresolved.
  This contrast of ease and difficulty gives the problem a classic
  charm.
\end{abstract}

\maketitle

To motivate the definition, we can start with the most elementary
example: the diagonals of a regular hexagon.  Any two of them cross
and define what the schoolbooks call supplementary vertical angles.
Without loss of information, we can take ``the'' angle defined by the
pair of lines to be the smaller of these two values.  Moreover, this
value is the same for all three possible pairs of lines: For any two
diagonals, their angle of intersection will be $\pi/3$.  We can state
this in a way amenable to generalization if we lay a unit vector along
each of the three diagonals.  Whichever way we choose to orient the
vectors, their inner products will satisfy
\begin{equation}
  |\langle v_j, v_k \rangle| = \left\{\begin{array}{cc}
  1, & j = k; \\
  \alpha, & j \neq k.
  \end{array}\right.
  \label{eq:definition}
\end{equation}
When a set of vectors $\{v_j:j=1\ldots,N\}$ enjoys this property, it
yields a set of equiangular lines.  This definition works equally well
in $\mathbb{R}^d$ and in $\mathbb{C}^d$.  An orthonormal basis is
equiangular, with $\alpha = 0$.  The question becomes more intriguing
when we push the size $N$ of the set beyond the dimension $d$.  It is
not terribly difficult to prove the \emph{Gerzon bound}: The size of a
set of equiangular lines cannot exceed $d(d+1)/2$ in $\mathbb{R}^d$ or
$d^2$ in $\mathbb{C}^d$~\cite{Lemmens:1973}. When the Gerzon bound is
met, the value of $\alpha$ is fixed, to $1/\sqrt{d+2}$ in
$\mathbb{R}^d$ and $1/\sqrt{d+1}$ in $\mathbb{C}^d$.  In the real
case, we know that we cannot in general attain the Gerzon bound, and
the question of how big $N$ can be as a function of $d$ ties in with
some of the most remarkable structures in discrete mathematics.
Meanwhile, in the complex case, it \emph{appears} that we can attain
the Gerzon bound in every dimension, but decades of work have not yet
settled the matter one way or the other --- and what we have seen so
far has already led us into deep questions of number theory and
quantum mechanics.

We take up the real case first.  In general, the discrete choice of
sign factors made when picking unit vectors to represent lines gives
the study of maximal equiangular line-sets in $\mathbb{R}^d$ a very
combinatorial flavor, and the theory of finite simple groups plays an
intriguing role.  A mere sampling of the known solutions and the
topics related to them will, regrettably, have to suffice.  The three
diagonals of a regular hexagon are a maximal set of equiangular lines
in $\mathbb{R}^2$, saturating the Gerzon bound.  Likewise, the six
diagonals of a regular icosahedron attain the Gerzon bound in
$\mathbb{R}^3$.  We start to fall short at dimension $d = 4$, where it
turns out we cannot exceed $N = 6$.  The only two other cases where
the Gerzon bound can be reached, as far as anyone knows, are in $d =
7$ and $d = 23$.  While the regular icosahedron dates back to the
semi-legendary centuries of ancient mathematics, thinking in terms of
maximal sets of equiangular lines is nowadays credited to Haantjes,
who solved the $\mathbb{R}^2$ and $\mathbb{R}^3$ cases in 1948. Van
Lint and Seidel resolved $\mathbb{R}^4$ through $\mathbb{R}^7$ in
1966~\cite{VanLint:1966}.  Even today, uncertainty starts to creep in
as soon as dimension $d = 17$~\cite{Greaves:2020}.  The Gerzon bound
can only be attained above $d = 3$ if $d + 2$ is the square of an odd
integer, but not all odd integers qualify: The maximum is known to be
strictly less than the Gerzon bound in dimensions 47 and
79~\cite{Barg:2014, Greaves:2016}.

But how remarkable those solutions in $d=7$ and $d=23$ are!  These
sets of equiangular lines can be extracted from celebrated structures
in one dimension higher, the $\mathrm{E}_8$ and Leech lattices
respectively.  The $N = 28$ lines in $\mathbb{R}^7$ are the diagonals
of the \emph{Gossett polytope} $3_{21}$, and they correspond among
other things to the 28 bitangents to a general plane
quartic~\cite{Manivel:2006}.  One way to obtain these lines --- there
are different constructions, but the results are all equivalent up to
an overall rotation --- stems from an observation by Van Lint and
Seidel~\cite{VanLint:1966, Seidel:1983} that if we want an interesting
set that involves the number seven, we ought to turn to the Fano plane
sooner or later. This geometry (the ``combinatorialist's coat of
arms'') is a set of seven points grouped into seven lines such that
each line contains three points and each point lies within three
distinct lines, with each pair of lines intersecting at a single
point. Consequently, if we take the \emph{incidence matrix} of the
Fano plane, writing a 1 in the $ij$-th entry if line $i$ contains
point $j$, then every two rows of the matrix have exactly the same
overlap:
\begin{equation}
  M = \begin{pmatrix}
     1 &1 &1 &0 &0 &0 &0 \\
     1 &0 &0 &1 &1 &0 &0 \\
     1 &0 &0 &0 &0 &1 &1 \\
     0 &1 &0 &1 &0 &1 &0 \\
     0 &1 &0 &0 &1 &0 &1 \\
     0 &0 &1 &1 &0 &0 &1 \\
     0 &0 &1 &0 &1 &1 &0
  \end{pmatrix}.
\end{equation}
The rows of the incidence matrix furnish us with seven equiangular
lines in~$\mathbb{R}^7$. To build this out into a full set of 28
lines, we can introduce sign factors~\cite{VanLint:1966, Seidel:1983},
and one way to do that is to add \emph{orientations} to the Fano
plane, exactly as one does when using it as a mnemonic for octonion
multiplication. We can label the seven points with the imaginary
octonions $e_1$ through $e_7$. When drawn on the page, a useful
presentation of the Fano plane has the point $e_4$ in the middle and,
reading clockwise, the points $e_1$, $e_7$, $e_2$, $e_5$, $e_3$ and
$e_6$ around it in a regular triangle. The three sides and three
altitudes of this triangle, along with the inscribed circle, provide
the seven Fano lines: $(e_1,e_2,e_3)$, $(e_1,e_4,e_5)$,
$(e_1,e_7,e_6)$, $(e_2,e_4,e_6)$, $(e_2,e_5,e_7)$, $(e_3,e_4,e_7)$,
$(e_3,e_6,e_5)$. The sign of a product depends upon the order, for
example, $e_1e_2 = e_3$ but $e_2 e_1 = -e_3$. The full multiplication
table, due to Cayley and Graves, is
\begin{equation}
\begin{array}{c|cccccccc}
e_i e_j & 1 & e_1 & e_2 & e_3 & e_4 & e_5 & e_6 & e_7 \\
\hline 
1 & 1 & e_1 & e_2 & e_3 & e_4 & e_5 & e_6 & e_7 \\
e_1 & e_1 & -1 & e_3 & -e_2 & e_5 & -e_4 & -e_7 & e_6 \\
e_2 & e_2 & -e_3 & -1 & e_1 & e_6 & e_7 & -e_4 & -e_5 \\
e_3 & e_3 & e_2 & -e_1 & -1 & e_7 & -e_6 & e_5 & -e_4 \\
e_4 & e_4 & -e_5 & -e_6 & -e_7 & -1 & e_1 & e_2 & e_3 \\
e_5 & e_5 & e_4 & -e_7 & e_6 & -e_1 & -1 & -e_3 & e_2 \\
e_6 & e_6 & e_7 & e_4 & -e_5 & -e_2 & e_3 & -1 & -e_1 \\
e_7 & e_7 & -e_6 & e_5 & e_4 & -e_3 & -e_2 & e_1 & -1
\end{array}
\end{equation}
which we can express visually by carefully assigning arrows to the
lines of the Fano plane.

To build our set of 28 equiangular vectors, start by taking the first
row of the incidence matrix $M$, which corresponds to the line $(e_1,
e_2, e_3)$, and give it all possible choices of sign by multiplying by
the elements not on that line. Multiplying by $e_4$, $e_5$, $e_6$ and
$e_7$ respectively, we get
\begin{equation}
  \begin{pmatrix}
 + &+ &+ &0 &0 &0 &0 \\
 - &+ &- &0 &0 &0 &0 \\
 - &- &+ &0 &0 &0 &0 \\
 + &- &- &0 &0 &0 &0
  \end{pmatrix}.
\end{equation}
Doing this with all seven lines of the Fano plane, we obtain a set of
28 vectors, each one given by a choice of a line and a point not on
that line. For any two vectors derived from the same Fano line, two of
the terms in the inner product will cancel, leaving an overlap of
magnitude 1. And for any two vectors derived from different Fano
lines, the overlap always has magnitude 1 because any two lines
always meet at exactly one point.

As we go up from $\mathbb{R}^7$ to $\mathbb{R}^{23}$, the properties
tap into more of the esoteric. As mentioned above, we can obtain the
$N = 276$ equiangular lines living in $\mathbb{R}^{23}$ from the Leech
lattice.  This lattice is how a grocer would stack 24-dimensional
oranges; the points of the lattice are the locations of the centers of
the spheres in the densest possible packing thereof in 24
dimensions~\cite{Cohn:2019}. The vectors from the origin to the
lattice points are classified by their ``type'', which is half their
norm.  To obtain a Gerzon-bound-saturating set of equiangular lines,
start with a vector of type 3.  For any such vector $\vec{v}$, there
are 276 unordered pairs of other lattice vectors having minimal norm
(type 2) that add to $\vec{v}$. Each pair specifies a line, and we
obtain 276 lines in all. These lines have rich group-theoretic
significance~\cite{Lemmens:1973, Taylor:1977}. First of all, their
symmetry group is Conway's third sporadic group $\mathrm{Co}_3$. The
\emph{stabilizer} of a line is the subgroup comprising those
symmetries of the set that leave that line fixed while permuting the
others.  For the $N = 276$ lines in $\mathbb{R}^{23}$, the stabilizer
of any line is isomorphic to the McLaughlin sporadic simple group
$\mathrm{McL}$.  Moreover, if we find the largest possible subset of
the $N = 276$ lines that are all orthogonal to a common vector in
$\mathbb{R}^{23}$, we obtain a set of 176 vectors that all squeeze
into $\mathbb{R}^{22}$, and these furnish a maximal set of equiangular
lines in that dimension.  The symmetries of this set constitute the
Higman--Sims sporadic simple group $\mathrm{HS}$. By close
consideration of the ways in which vector directions can be assigned
to the $N = 276$ lines, it is also possible to distinguish special
subsets of $23$ lines and thence obtain the Mathieu group
$\mathrm{M}_{23}$~\cite{Gillespie:2018}.

\begin{table}
\begin{tabular}{cccccccccc}
  $d$ & \textbf{2} & \textbf{3} & 4 & 5 & 6 & \textbf{7}--14 & 15 & 16 & 17\\
$N_{\rm max}(d)$ & 3 & 6 & 6 & 10 & 16 & 28 & 36 & 40 &
  48--49 \\
$d$  & 18 & 19 & 20 & 21 & 22
& \textbf{23}--41 & 42 & 43 \\
$N_{\rm max}(d)$  & 56--60 & 72--75 & 90--95 & 126 & 176 & 276 & 276--288 & 344
\end{tabular}
\caption{\label{tab:bounds} Bounds on the largest possible size of a
  set of equiangular lines in $\mathbb{R}^d$.  For some values of the
  dimension $d$, the bound is not known exactly.  Dimensions where the
  Gerzon bound is saturated are shown in bold.  For more details, see
  \texttt{OEIS:A002853} and references therein.}
\end{table}

We do not yet have a general theory of how the maximum $N$ varies with
the dimension $d$, and of course, we can only hazard guesses about
what the textbooks of tomorrow might contain, but one conceptual
connection that has proved quite important so far is to algebraic
graph theory.  If $\{v_i : i=1,\ldots,N\}$ is a set of unit vectors
defining a set of equiangular lines, then by the definition of
equiangularity, the Gram matrix of these unit vectors will be 1 along
the diagonal and $\pm \alpha$ everywhere else.  Note that specifying a
choice of sign for each off-diagonal entry is exactly the same
information needed to specify which edges are connected in a graph
with $N$ vertices.  Changing the sign of any vector will flip the
signs on some of the entries in the Gram matrix, which rewires the
corresponding graph in a particular way.  So, a set of $N$ equiangular
lines corresponds to an equivalence class of $N$-vertex graphs (a
``switching class''), and there is a natural overlap of interest
between equiangular lines and strongly regular
graphs~\cite{Cameron:2004, Gillespie:2018, Greaves:2016, Sustik:2007}.

Because the matrices one tries to construct when attempting to build
sets of real equiangular lines will be filled with integers,
conditions for their existence will be equations that algebraic number
theory is suited to handle. In particular, the study of cyclotomic
fields --- fields made by extending $\mathbb{Q}$ with a primitive
$n$th root of unity --- becomes relevant~\cite{Sustik:2007}.

One fruitful avenue of inquiry has been to \emph{fix an angle} and ask
how many lines in $\mathbb{R}^d$ can be equiangular with that chosen
angle~\cite{Barg:2014, Greaves:2016, King:2019}.  Balla et al.\ have
proved that for a fixed $\alpha$, when the dimension $d$ becomes
sufficiently large there can be at most $2(d-1)$ lines in
$\mathbb{R}^d$ that are equiangular with common overlap
$\alpha$~\cite{Balla:2018}.

For students who wish to jump in and do calculations as quickly as
possible, Tremain~\cite{Tremain:2008} provides a useful collection of
constructions.

The complex case was first studied as a natural counterpart to the
real one.  Investigations of structures like complex
polytopes~\cite{Coxeter:1991} turned up maximal sets of complex
equiangular lines in dimensions $d = 2$, $3$ and $8$.  Then the 1999
PhD thesis of Zauner~\cite{Zauner:1999}, followed by the independent
work of Renes et al.~\cite{Renes:2004}, made complex equiangular lines
into a physics problem.  Now we have exact solutions for 102 different
values of the dimension, including all dimensions from 2 through 40
inclusive, and some as large as $d = 1299$.  Furthermore, numerical
solutions are known to high precision for all dimensions $d \leq 193$,
and some as large as $d = 2208$~\cite{Bengtsson:2020,
  Horodecki:2020}. These lists have grown irregularly, since different
simplifications have proved applicable in different dimensions. (We
endeavor to keep the website~\cite{UMB} up to date.)  Credit is due to
M.\ Appleby, I.\ Bengtsson, T.-Y.\ Chien, S.\ T.\ Flammia, M.\ Grassl,
G.\ S.\ Kopp, A.\ J.\ Scott and S.\ Waldron.

Physicists know sets of $d^2$ equiangular lines in $\mathbb{C}^d$ as
``Symmetric Informationally Complete Positive-Operator-Valued
Measures'', a mouthful that is abbreviated to \emph{SIC-POVM} and
often further just to \emph{SIC} (pronounced ``seek'').  The
``Symmetric'' refers to the equiangularity property, while the rest
summarizes the role these structures play in quantum
theory~\cite{Bengtsson:2017, Waldron:2018}.  A basic premise of
quantum physics is that to each physical system of interest is
associated a complex Hilbert space $\mathcal{H}$.  The subdisciplines
of quantum information and computation~\cite{Nielsen:2010} often
employ finite-dimensional Hilbert spaces, $\mathcal{H}_d \simeq
\mathbb{C}^d$.  The mathematical representation of a \emph{measurement
  process} is a set of positive semidefinite operators $\{E_j\}$ on
$\mathcal{H}_d$ that sum to the identity.  Each operator in the set
stands for a possible outcome of the measurement, and the set as a
whole is known as a POVM.  To represent the preparation of a quantum
system, we ascribe to the system a \emph{density operator} $\rho$ that
is also a positive semidefinite operator on $\mathcal{H}_d$, in this
case normalized so that its trace is unity.  The \emph{Born rule}
states that the probability for obtaining an outcome of a measurement
is given by the Hilbert--Schmidt inner product of the density operator
and the POVM element representing that outcome:
\begin{equation}
  p(j) = \mathrm{tr}\, \rho E_j.
\end{equation}
If the POVM $\{E_j\}$ has at least $d^2$ elements, then it is possible
for it to span the space of Hermitian operators on $\mathcal{H}_d$.  In this
case, the POVM is \emph{informationally complete} (IC), because any
density operator can be expressed as its inner products with the POVM
elements. Or, in more physical terms, the probabilities for the
possible outcomes of an IC POVM completely specify the preparation of
the quantum system.  Given a set of unit vectors in $\mathbb{C}^d$
defining $d^2$ equiangular lines, the projectors $\{\Pi_j\}$ onto
these lines span the Hermitian operator space, and up to normalization
they provide a resolution of the identity:
\begin{equation}
  \sum_{j=1}^{d^2} \Pi_j = dI.
\end{equation}

SICs satisfy a host of optimality conditions. By many standards, a SIC
is as close as one can possibly get to having an orthonormal basis
for Hermitian operator space while staying within the positive
semidefinite cone.  This is highly significant for quantum theory,
because positivity is a crucial aspect of having operator
manipulations yield well-defined probabilities.

Decades of work on the fundamentals of quantum physics have shown that
quantum uncertainty cannot be explained away as ignorance about
``hidden variables'' intrinsic to physical systems but concealed from
our view.  Historically, this area has been home to deep theorems like
the results of Bell, Kochen and Specker~\cite{Mermin:1993}, while in
the modern age, it is also a topic of increasingly practical
relevance, since if we want our quantum computers to be worth the
expense, we had better understand exactly which phenomena can be
imitated classically and which cannot~\cite{Veitch:2014}.  SICs
provide a new window on these questions by furnishing a measure of the
margin by which any attempt to model quantum phenomena with intrinsic
hidden variables is guaranteed to fail~\cite{DeBrota:2018}.

Here we have a peculiar confluence of topics advanced by the late John
Conway.  It was his insight into the Leech lattice that gave us the
maximal equiangular set in $\mathbb{R}^{23}$, and together with Simon
Kochen he carried forward the study of how hidden-variable hypotheses
break down~\cite{Conway:2006}. Somewhere in the world's weight of loss
is the fact that we will never know what he might have thought about
these two problems coming together.

Above, we noted the group-theoretic properties of real equiangular
lines.  Group theory also manifests in the complex case.  First, all
known SICs are \emph{group covariant}, meaning that they can be
generated by taking a well-chosen initial vector and computing its
orbit under a group action.  Moreover, in all cases except a class of
solutions in $d = 8$, the group in question is the
\emph{Weyl--Heisenberg group} for dimension $d$.  Given an orthonormal
basis $\{e_n:i=1,\ldots,d\}$, we define a shift operator $X$ that
sends $e_n$ to $e_{n+1}$ modulo $d$, and a ``clock'' or ``phase''
operator $Z$ that sends $e_n$ to $\exp(2\pi in/d) e_n$.  The two
unitary operators $X$ and $Z$ commute up to the phase factor
$\exp(2\pi i/d)$, and together with phase factors their products
define the Weyl--Heisenberg group.  (The $d = 2$ special case is also
known as the Pauli group.  Weyl introduced $X$ and $Z$ in 1925, in
order to define what the quantum mechanics of a discrete degree of
freedom could mean~\cite{Weyl:1931}.  The association of Heisenberg's
name with this group is a convention that has less historical
justification, since the ``canonical commutation relation'' of
position and momentum that inspired Weyl was not due to
Heisenberg~\cite{Born:2002, Fedak:2009}.)  The Weyl--Heisenberg group
is significant in multiple aspects of quantum theory, such as the
study of when quantum computations can be efficiently emulated
classically.  Zhu has proved that when the dimension $d$ is prime,
group covariance implies Weyl--Heisenberg covariance~\cite{Zhu:2012}.
However, it is not known whether SICs must be group covariant in
general.

In dimension $d = 8$, there exist Weyl--Heisenberg SICs but also a
class of SICs covariant under a different group (the tensor product of
three copies of the Pauli group).  These were first discovered by
Hoggar and can be termed the \emph{Hoggar-type}
SICs~\cite{Hoggar:1981, Hoggar:1998, Szymusiak:2015}.  All of them are
equivalent to one another under unitary and anti-unitary conjugations.
Zhu discovered that the stabilizer of an element in a Hoggar-type SIC
is always isomorphic to $\mathrm{PSU}(3,3)$, the group of $3 \times 3$
unitary matrices over the finite field of order 9~\cite{Zhu:2015a}.
This is moreover the commutator subgroup of the automorphism group of
the Cayley integral octonions~\cite{Wilson:2009}, also known as the
\emph{octavians}~\cite{Conway:2003}.  In other words, the linear maps
from the octavians to themselves that preserve the multiplication
structure form a group, and an index-2 subgroup of that gives the ways
to hold one vector in a Hoggar-type SIC in place and permute the other
63.

The octavians form a lattice, and up to an overall scaling, it is the
same as the $\mathrm{E}_8$ lattice.  So, a symmetric arrangement of
\emph{complex lines} is, under the surface, tied in with an optimal
packing of \emph{real hyperspheres}~\cite{Viazovska:2017} --- a
development that was completely unforeseen.

Exploration of the Weyl--Heisenberg SICs leads into algebraic number
theory, and at a more subtle and demanding level than that subject has
so far manifested in the study of real equiangular lines.  During the
early years, SIC vectors were found by computer algebra, laborious
calculations with Gr\"obner bases yielding coefficients that were, in
technical terms, ghastly~\cite{Scott:2010a}.  Pages upon pages of
printouts were sacrificed.  Yet, upon closer study, these solutions
turned out to be not as bad as they seemingly could have been: In $d
\geq 4$, they were always expressible in terms of nested radicals.
Since the polynomial equations being solved were of degrees much
higher than quintic, there was no reason to expect a solution by
radicals, and so the Galois theory of the SIC problem became a topic
of interest~\cite{Appleby:2013}.  This has led to a series of
surprises.

Recall that a group $G$ is solvable if we can write a series starting
with the trivial group, $\langle 1\rangle < H_1 < \cdots < H_m < G$,
where each of the $H_i$ is a normal subgroup of the next and the
quotients $H_i/H_{i-1}$ are all abelian.  The zeros of a polynomial
$f(x)$ over some field $\mathbb{K}$ can be found by radicals exactly
when the splitting field of $f$, the extension of $\mathbb{K}$ in
which $f$ falls apart into linear factors, can be made by stacking up
abelian extensions of $\mathbb{K}$.  So, if we want to understand
solvability by radicals, getting a handle on abelian extensions of
number fields is the thing to do.  One obvious place to start is
abelian extensions of the rationals $\mathbb{Q}$, and the
\emph{Kronecker--Weber theorem} tells us that every abelian extension
of $\mathbb{Q}$ is contained in a cyclotomic field.

Hilbert's twelfth problem asks for a broadening of the
Kronecker--Weber theorem, or in other words, a classification of the
abelian extensions of arbitrary number fields.  This problem remains
unresolved, although progress has been made.  When we generalize
beyond the case where the base field is $\mathbb{Q}$, the role of the
cyclotomic fields is played by the \emph{ray class fields.}  The
generalization of the $n$ in a cyclotomic field's $n$th root of unity
is a number called the \emph{conductor} of the ray class field.  In
the original case covered by Kronecker--Weber, the fields in which the
abelian extensions all live are generated by special values of a
special function, i.e., the exponential function evaluated at certain
points along the imaginary axis.  What functions play the role of the
exponential more generally, and at what points should they be
evaluated?  This is much more difficult to say.

Historically, the first to be understood beyond abelian extensions of
the rationals themselves were abelian extensions of imaginary
quadratic fields, that is, $\mathbb{Q}(\sqrt{-n})$ where $n$ is a
positive integer.  This is significantly more demanding than the case
where the base field is $\mathbb{Q}$.  The theory that does the job
goes by the name \emph{complex multiplication}, which to a mathematics
student is deceptively simple.  (Perhaps a term like ``elliptic
multiplication'' would be better, but here as elsewhere, the jargon
has solidified.) This theory is informally described as an order of
magnitude harder than the Kronecker--Weber theorem, and the case of
\emph{real} quadratic fields is more difficult still, and only
partially understood.

SICs know a lot about Hilbert's twelfth problem~\cite{Appleby:2017b}.
They have been found to generate ray class fields over real quadratic
extensions of the rationals $\mathbb{Q}(\sqrt{D})$, where $D$ is the
square-free part of the quantity $(d-3)(d+1)$.  The path to
counterparts for roots of unity goes through the \emph{overlap
  phases,} the phases that the absolute-value bars in
Eq.~(\ref{eq:definition}) throw away.  The overlap phases turn out to
be \emph{algebraic integer units} in ray class fields or extensions
thereof. (Roots of monic polynomials over $\mathbb{Z}$ are the
algebraic integers, so called because their quotients yield the
algebraic numbers just as quotients of ordinary integers $\mathbb{Z}$
yield the rationals $\mathbb{Q}$. The algebraic integers within a
number field form a ring, and the units of this ring are those
algebraic integers whose multiplicative inverses are also algebraic
integers. In the real case, taking the absolute value discards a
choice of $\pm 1$, while in the complex case, it discards a phase that
is not arbitrary, but rather a generalized kind of ``$\pm 1$''!)
Recently, Kopp has turned this connection around and, using
conjectured properties of important numbers in ray class fields,
constructed an exact SIC in $d = 23$ where none had been known
before~\cite{Kopp:2018}.  Kopp's SIC is constructed from overlap
phases calculated as Galois conjugates of square roots of \emph{Stark
  units}.  These numbers figure largely in the Stark conjectures,
which pertain to generating ray class fields explicitly.  The
conceptual waters here run deep, yet more remarkable still is the fact
that these beguiling questions of number theory are, by way of almost
schoolchildish geometry, intricated with quantum physics.

\bibliographystyle{utphys}
\bibliography{what-is-equiangular}

\end{document}